\documentclass[12pt,a4paper]{article}

\usepackage{amsmath}
\usepackage{amsfonts}
\usepackage{amssymb}
\usepackage{amsthm}

\input{epsf}             
\def\epsfcenter#1{{\vcenter{\hbox{\epsfbox{#1}}}}} 
\def\epsfpad#1#2#3{{\vcenter{\vskip#2\hbox{\epsfbox{#1}}\vskip#3}}} 

\def\epsfsize#1#2{0.5#1} 

\newcommand{\rd}{\mathrm{d}}            
\newcommand{\R}{\mathbb R}              
\newcommand{\Z}{\mathbb Z}              

\newcommand{\captionnotext}{\refstepcounter{figure}\begin{center}Figure \thefigure\end{center} }  

\newcommand{\fig}[1]{Figure \ref{#1}}  

\theoremstyle{definition}
\newtheorem*{theorem}{Theorem} 

\begin{document}

\title{Geometrical measurements in three-dimensional quantum gravity}

\author{John W. Barrett
\thanks{This is the first of three lectures given at the Xth Oporto Meeting on Geometry, Topology and Physics, September 2001. Copyright \copyright\ John W. Barrett 2002}
\\ \\
School of Mathematical Sciences\\
University of Nottingham\\
University Park\\
Nottingham NG7 2RD, UK\\
\\
E-mail john.barrett@nottingham.ac.uk}

\date{February 14th, 2005}

\maketitle

\begin{abstract} A set of observables is described for the topological
quantum field theory which describes quantum gravity in three space-time
dimensions with positive signature and positive cosmological constant. The
simplest examples measure the distances between points, giving spectra and
probabilities which have a geometrical interpretation. The observables are
related to the evaluation of relativistic spin networks by a Fourier
transform. 
\end{abstract}


\section{Distances}
\begin{figure}[ht]
\begin{center}\ \epsfbox{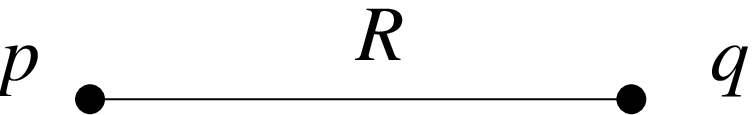} \end{center}
\captionnotext \label{geodesic}
\end{figure}

In general relativity we can measure the distance $R$ between a pair of points by
considering the length of a geodesic between them (\fig{geodesic}).

In quantum gravity the metric fluctuates, so we expect only to be able to say what
the possible values for $R$ are, and their probabilities. In general we might expect these to depend on the topology of the manifold $M$ in which the points $p$ and $q$ lie.

The Turaev--Viro state sum model~\cite{TV} gives a theory of quantum gravity in 3 dimensions
where the metric has positive signature~\cite{QGTQFT}; it gives a concrete method for calculating the functional integral for three-dimensional gravity~\cite{WI3D}. You can think of this as related to a 4-dimensional theory with $-+++$ signature metrics where the time dimension ($-$) has been dropped. Although this is not entirely realistic, it does give us a model in which the relationship between classical geometry and quantum gravity can be explored.

The model is specified by an integer
$r\geq 3$. Given points $p,q \in M$ connected by a curve then an observable
can be defined which takes values in the set of {\bf spins}
\[
j\in \left\{ 0, \frac{1}{2},1, \dots , \frac{r-2}{2}\right\}.
\]
The probability that the spin takes value $j$ can be calculated to be
\[
P_j=\frac{(\dim _q j)^2}{N},
\]
with $\dim _q j$ the quantum dimension of the spin $j$ representation of $U_qsl(2)$
for $q=e^{i\pi/r}$, and $N$ a normalisation constant.  The formula for the quantum dimension is
\[
\dim _q j=(-1)^{2j} \left( \frac{\sin \frac{\pi}{r}(2j+1)}{\sin \frac{\pi}{r} } \right)
\]
and $N=\sum _j (\dim _q j)^2$ is the constant which ensures that $\sum _j P_j=1$. Actually in this model the distance measurements only depend on the topology of $p$, $q$ and $M$ to the extent that the points $p$ and $q$ are required to be in the same connected component of $M$. The topology of $M$ comes into generalizations of the formula considered further below. First I will describe how the probability formula is calculated, and then its physical interpretation. 
 
 \pagebreak[4] 
 
\section{Calculation}

The Turaev-Viro state sum for a closed compact manifold $M$ is a formula for an invariant $Z(M)\in \R$. This is defined with the aid of a triangulation of the manifold; however the value of $Z(M)$ is independent of the triangulation chosen and depends only on the topology of $M$. 

 \begin{figure}[ht]
\begin{center}\ \epsfbox{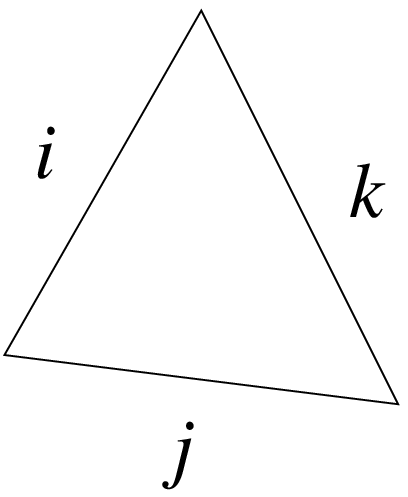} \end{center}
\caption{State for a triangle} \label{trianglestate}  
\end{figure}

A state for this state sum model is the assignment of a spin $i,j,k,\ldots$ to each edge of the
triangulation (\fig{trianglestate}) such that the following `admissibility' conditions for each triangle are satisfied.

\begin{gather}
i\leq j+k\label{trieq1}\\
j\leq k+i\label{trieq2}\\
k\leq i+j\label{trieq3}\\
\label{spheq}
i+j+k\leq r-2\\
\label{parityeq}
i+j+k=0 \quad \mod 1
 \end{gather}

Given a state, each simplex is assigned a weight, a real number. This number  depends on the spin labels for the edges in that simplex. The weights are calculated using the spin network evaluation based on the Kauffman bracket $\left<\mathstrut\quad\right>$ with $A=e^{i\pi r/2}$ as follows \cite{KB}:
\begin{center}
\begin{tabular}{|c|c|c|}
\hline
Simplex &\quad\quad Weight \quad\quad&Spin Network\\
\hline
$\epsfpad{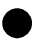}{10pt}{10pt}$&
$N^{-1}$&\\
$\epsfpad{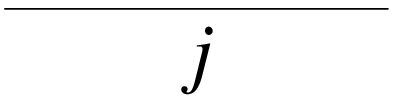}{10pt}{10pt}$&
$\dim _qj$ & $\dim _qj=\left<\epsfpad{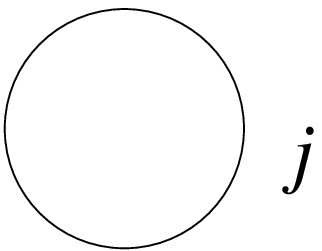}{5pt}{5pt}\right>$\\
$\epsfpad{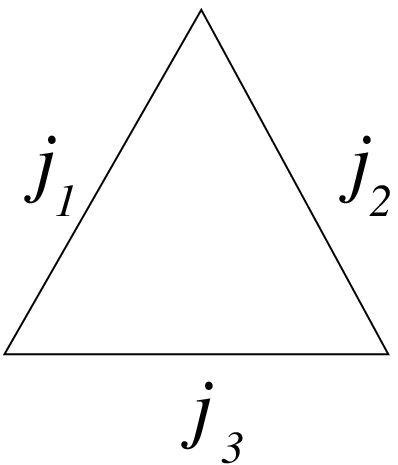}{5pt}{5pt}$ &
$\Theta^{-1}$ & 
$\Theta=\left<\epsfpad{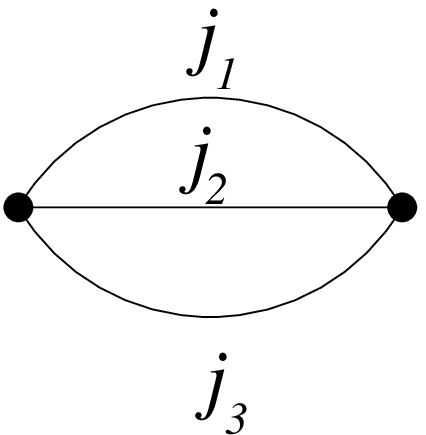}{5pt}{5pt}\right>$\\
$\epsfpad{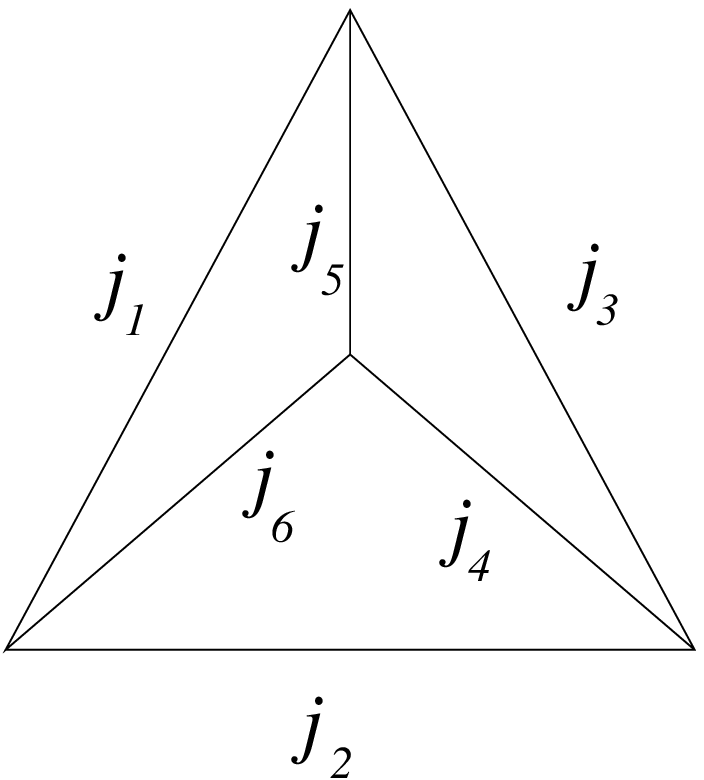}{5pt}{5pt}$ &
$\tau$ &
$\tau=\left<\epsfpad{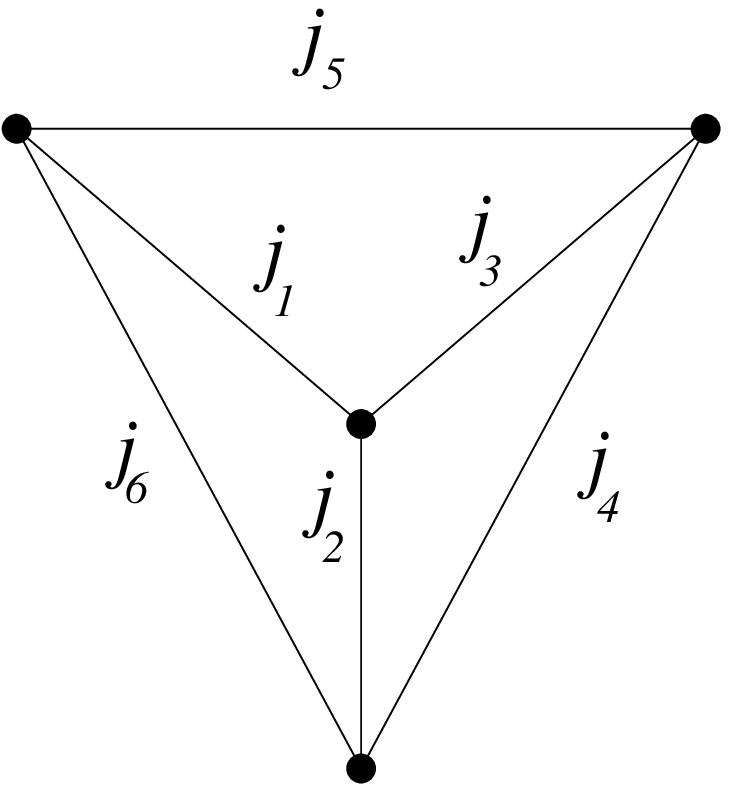}{5pt}{5pt}\right>$\\
\hline
\end{tabular}
\end{center}
 For the tetrahedral simplex, the spin network in the right-hand column is the graph which is dual to the edges of the tetrahedron.

The state sum formula is
\[
Z(M)=\sum _{ {\text{states}}}\; \prod _{\text{simplexes}}
\text{weights}.
\]

The probability formula is calculated by finding a triangulation
of $M$ such that $p$ and $q$ are the two vertices of a single edge in the triangulation.
 The probability $P_j$ is just the probability that the distinguished edge has its spin equal to $j$, i.e.
\[
P_j=\frac{Z(M,j)}{Z(M)},
\]
where $Z(M,j)$ is the sum over the subset of states that have spin $j$ on the distinguished edge. Clearly $Z(M)=\sum _j Z(M,j)$, so the $P_j$ sum to $1$. 

The proof that the formula for $P_j$ is correct is
to calculate it explicitly for a particular triangulation, and then use the triangulation invariance of the  state sum formula to show that the
formula holds for all triangulations which have an edge that runs from $p$ to $q$. The calculation can be done easily for $M=S^3$ using the singular triangulation of $S^3$ with two tetrahedra. It also follows from the Fourier transform result proved below. The proof that the result is the same for any triangulation will appear elsewhere. The fact that the answer is the same for any manifold follows from the connected sum formula for the Turaev-Viro invariant and the fact that the edge is contained in a ball in $M$.

The positivity of the probabilities is not immediately obvious from the definition of the state sum since the total weight for a state of $M$ can have either sign. However it does follow from the fact that the Turaev-Viro model has state spaces on surfaces which are Hilbert spaces, and  the state sum formula for $P_j$ is the expectation value of a positive operator (a projector) on the Hilbert space of $S^2$.

\section{Geometrical Models} A physical interpretation is most apparent
in the limiting case $r\to \infty$ (the Ponzano--Regge model~\cite{PR}).  Then, 
\[
P_j=\frac{(2j+1)^2}{N}. 
\] 
There is however no value for $N$ which
normalises $\sum P_j$ to 1 so this limit is somewhat degenerate. 
Nevertheless, Ponzano and Regge discovered that the asymptotic formulae
for the state sum in the limit $j\to \infty$ have a geometric
interpretation if one takes $j+\frac{1}{2}$ to be the length of the edge
in 3-dimensional Euclidean space $\mathbb{R}^3$.  Also, they suggested
that the semi-classical configurations of the state sum model are given
by mapping the simplicial complex to $\mathbb{R}^3$ with an approximate
uniform measure for the position of the vertices in $\mathbb{R}^3$. This
is also consistent with the gauge theory interpretation of the model in
which the gauge group is the semi-direct product of $SU(2)$ and $\R^3$~\cite{WITCA}.

These considerations suggest that for spin $j$, the distance between $p$
and $q$ is $j+\frac{1}{2}$ and the probability $P_j$ is proportional to
the area of the 2-sphere of radius $j+\frac{1}{2}$.  In other words, a
geometric model for the probabilities $P_j$ is to consider displacement
vectors in $\mathbb{R}^3$ which have length $R=j+\frac{1}{2}$ but
undetermined direction.  The measure $P_j$ is a discrete version of the
uniform measure $4 \pi R^2 dR$ in three-dimensional Euclidean space. This
gives a model for the $P_j$ in terms of probability measures for points
moving in the classical geometry.

\begin{figure}[ht]
\begin{center}\epsfbox{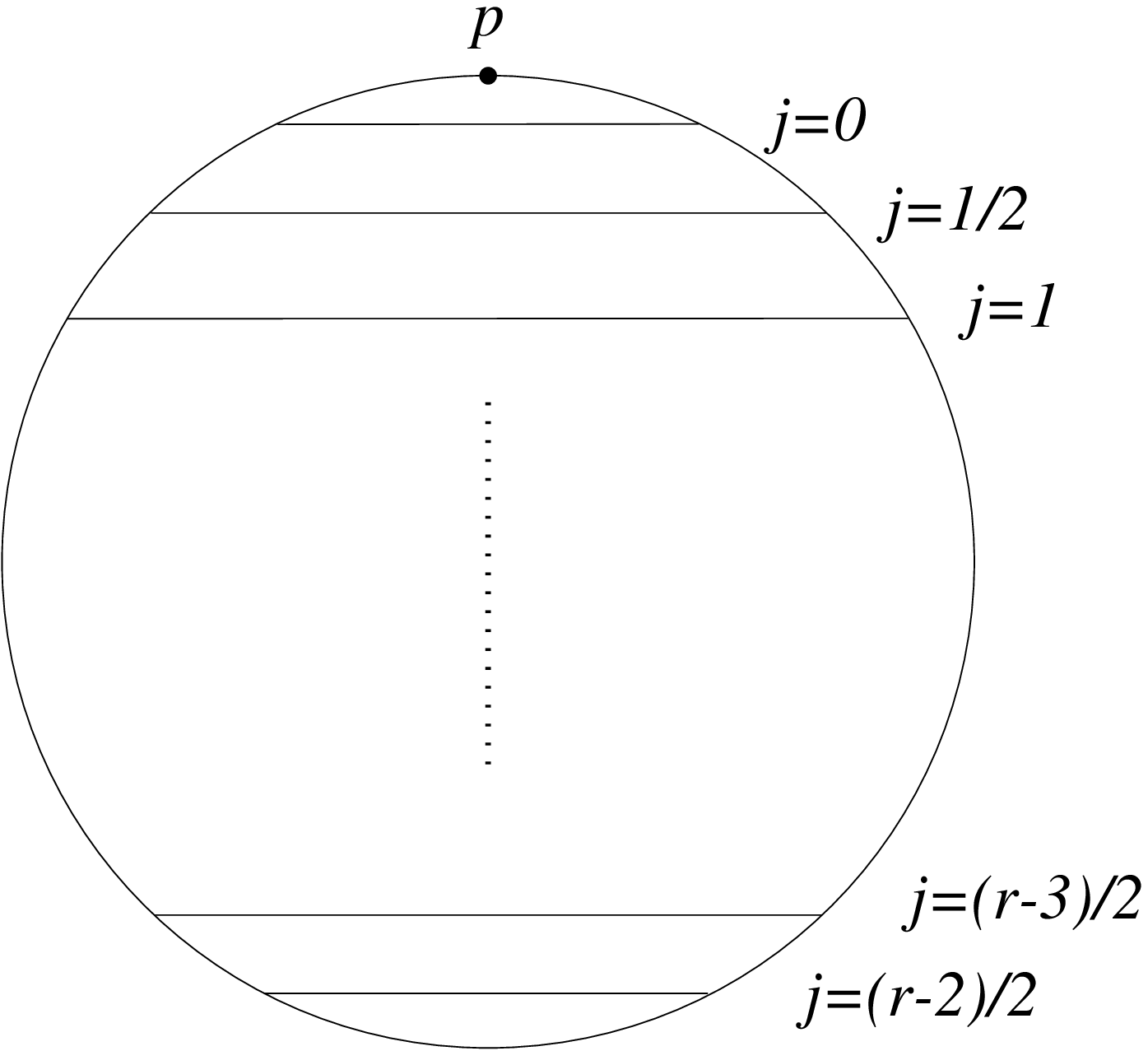} \end{center}
\caption{Possible orbits for $q$} \label{sphere}
\end{figure}

Now to return to the Turaev--Viro model.  The Lagrangian quantum field theory view is
that this model is a version of quantum gravity with a positive cosmological constant
$\Lambda$, whereas the Ponzano--Regge model has $\Lambda =0$.  The classical solutions
are locally a 3-sphere, with radius $\sqrt{1/\Lambda}$.  Obviously as
$\Lambda \to \infty$ this degenerates to the Euclidean space $\mathbb{R}^3$ of the Ponzano-Regge model.  This
suggests that the physical interpretation of the probabilities $P_j$  should be based on configurations in $S^3$.  Indeed the area of a 2-sphere of radius $j+\frac{1}{2}$ in $S^3$ is
\begin{equation}
\text{area}=4\pi \sin ^2 \frac{\pi}{r} (2j+1), \label{area}
\end{equation}
i.e., proportional to $P_j$, if the 3-sphere has radius $r/{2\pi}$.  If the
point $p$ is fixed at the `north pole' then the possible positions for $q$ lie on the
2-spheres indicated on \fig{sphere}
with probability proportional to the area. In the figure, the 3-sphere is projected to a disk on the plane and the 2-sphere of constant height is shown in its projection as a horizontal line.
 
 In this way the range of values for the
spin also has a natural explanation in terms of the 3-sphere: the lengths take all possible half-integer values for distances on the 3-sphere of radius $r/2\pi$. This only works because of the `$+\frac{1}{2}$' in the relation between spin and distance. The minimum distance is then $1/2$ and the maximum $(r-1)/2$. There are
two other possible half-integral values for the distance between a pair of points,
namely 0 and the half-circumference $r/2$.  However the corresponding probabilities in this picture
are zero, and so these possibilities don't occur.

\section{Generalizations}
\begin{figure}[ht]
\begin{center}\epsfbox{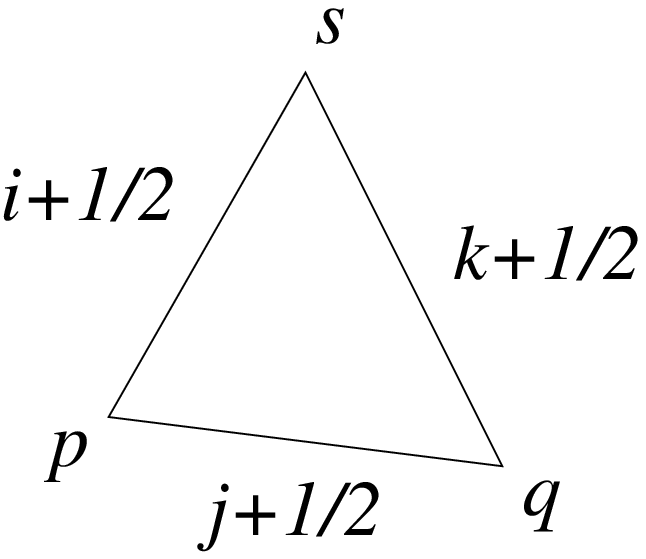} \end{center}
\caption{Geometry for a triangle}\label{trianglegeom}
\end{figure}

In a similar way we can calculate the probability in the state sum model for three points $p,q,s$ which are
the vertices of an embedded triangle in $M$ to be separated by distances $i+\frac{1}{2}$,
$j+\frac{1}{2}$, $k+\frac{1}{2}$ (\fig{trianglegeom}).
The result is
\[
P(i,j,k)=\frac{Z(M;i,j,k)}{Z(M)}=\left\{
\begin{array}{ll}
N^{-2}\dim _qi \ \dim _q j \ \dim _q k&\text{if } (i,j,k)\ \text{admissible},\\
0&\text{else}.
\end{array}
\right.
\]
In carrying out this calculation, the topological configuration is important. One has to specify curves which connect each pair of points. What is important is that the loop of the three edges is unknotted and is a contractible loop in $M$, in other words that the three edges do indeed bound a triangle in $M$.

\begin{figure}[ht]
\begin{center}\epsfbox{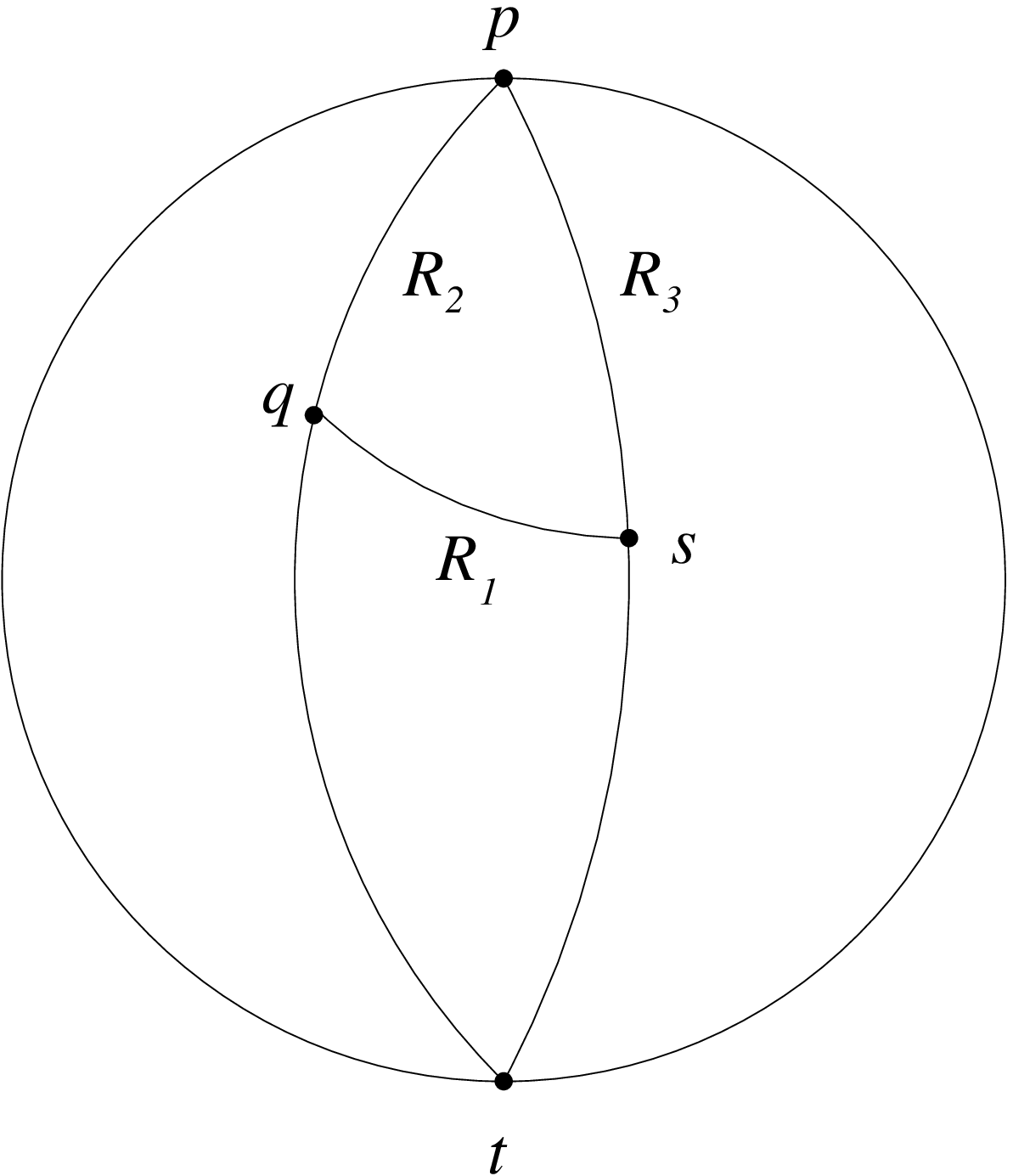} \end{center}
\caption{Geodesic triangle on a sphere} \label{proof}    
\end{figure}

The non-zero part of this formula is 
\begin{equation} \sin \frac \pi r (2i+1) \; \sin \frac \pi r (2j+1)\; \sin \frac \pi r (2k+1) \label{triangle}
\end{equation}
which is positive, the signs cancelling due to the admissibility condition (\ref{parityeq}). The other four admissibility conditions (\ref{trieq1})--(\ref{spheq}) have a geometrical interpretation when they are rewritten in terms of the lengths:
\begin{gather}
i+\frac{1}{2}<(j+\frac{1}{2})+(k+\frac{1}{2})\\
j+\frac{1}{2}<(k+\frac{1}{2})+(i+\frac{1}{2})\\
k+\frac{1}{2}<(i+\frac{1}{2})+(j+\frac{1}{2})\\
(i+\frac{1}{2})+(j+\frac{1}{2})+(k+\frac{1}{2})<r\label{spheq2}
\end{gather}
The first three are interpreted as the conditions for the edge-lengths of a non-degenerate triangle in a metric space geometry.  A triangle is degenerate if there is a vertex whose location is  uniquely determined by the location of the other two vertices.
However the fourth condition is again specific to a sphere: a geodesic triangle with sides $R_1,R_2,R_3$ on a sphere (in any dimension) of radius $r/2 \pi$ satisfies
the inequality $R_1+R_2+R_3\le r$.
The proof of this is very simple. The triangle inequalities for $tqs$ (\fig{proof}) give
\[
R_1\leq (r/2-R_2)+(r/2-R_3) \qquad \text{or} \qquad R_1+R_2+R_3\leq r.
\]

However in the case $R_1+R_2+R_3=r$ the three points lie on a diameter and one of the points is determined uniquely by the location of other two. Such a triangle is therefore degenerate. The overall result is that the conditions (\ref{trieq1}-\ref{spheq}) are the conditions for a non-degenerate triangle on $S^3$.

The geometrical model (\ref{area}) for the single edge can be extended to this case. Consider three points $p$, $q$ and $s$ on $S^3$ with a uniform probability distribution. The probability that the distances between them are $R_1$, $R_2$ and $R_3$, as in \fig{proof}, is proportional to
$$ \sin \frac {2\pi R_1} r \;\sin \frac {2\pi R_2}r \;\sin \frac {2\pi R_3}r  \; \rd R_1\; \rd R_2 \;\rd R_3$$
as long as the inequalities for a triangle are satisfied (Appendix 1).
This formula is the continuum analogue of (\ref{triangle}), and in fact (\ref{triangle}) is obtained by substituting $R_1=i+1/2$,  $R_2=j+1/2$, $R_3=k+1/2$ in this probability density. This means that the geometrical model reproduces the measure $P(i,j,k)$ under the additional assumption that all edge length are required to be a half-integer.

In a similar way one can analyse an embedded polygon in $M$, obtaining
probabilities which can be considered as a measure of the volume of
configurations of an unknotted circular loop of rods of fixed length in
$S^3$.  It is an interesting problem to relate this to other measures of
the volume of these configurations, such as the symplectic volume measure
provided in the flat ($r\to \infty$) case by the Riemann--Roch theorem~\cite{HK, KM}.

  These simple examples may give the misleading impression that the classical geometry is always the standard metric 3-sphere. However this is not the case, as the observable is sensitive to knotting and linking. The general situation is studied in the next section.

\section{Fourier transform}
  In general one can consider the set of edges on which the
spins are fixed to form an embedded graph $\Gamma$ in $M$.  Then the state sum invariant with these spins fixed gives an invariant of the embedded graph under motions of the graph in the manifold (ambient isotopies).\footnote{ A different set of observables to the ones investigated here were defined in \cite{TBIG, KS}.}

In the case of $M=S^3$ there is another invariant of embedded graphs with edges labelled by spins, the relativistic spin network invariant defined by Yetter~\cite{YGR,CLEV}. In this section it is shown that the two invariants are related by a Fourier transform of the spin labels. This substantially generalises the $\Z_2$ Fourier transform of \cite{ROEX,YEX}.

The definition of the relativistic spin network invariant is as follows.  Let $\Gamma(i_1,i_2,\ldots,i_n)$
be a  graph embedded in $S^3$, and its edges labelled with spins $i_1,i_2,\ldots,i_n$ (in a fixed order). First, the invariant is defined in the case of trivalent graphs, then this will be generalised to arbitrary vertices.
For each vertex of a trivalent graph there are three spin labels $(i,j,k)$ on the three edges meeting the vertex. The invariant is defined to be zero unless each triple satisfies the admissibility conditions  (\ref{trieq1})--(\ref{parityeq}). Suppose that these conditions are satisfied for each vertex.
Put 
$$\Theta=\left<\epsfcenter{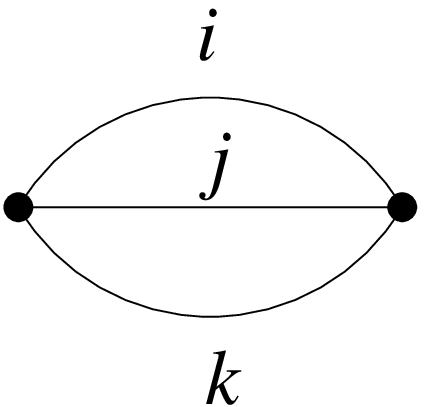}\right>$$
Then the relativistic invariant $\left\langle \Gamma(i_1,i_2,\ldots,i_n) \right\rangle_R$ is defined in terms of
the Kauffman bracket invariant of the diagram given
by projecting the graph in $S^3$ to $S^2$ by
\[
\left\langle \Gamma \right\rangle_R = 
\frac{\left| \left\langle \Gamma \right\rangle \right| ^2} {\prod_{\text{vertices}}\Theta}.
\]
This definition is extended to arbitrary graphs by the relations
\begin{equation*}
\left\langle \epsfcenter{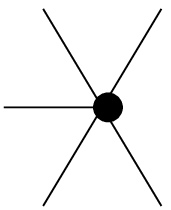}\,\right\rangle _R=\sum _j \left\langle \epsfcenter{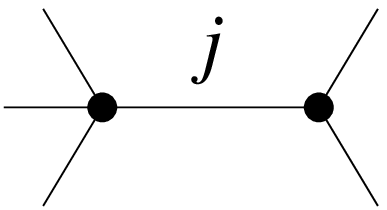} \,\right\rangle _R \
\dim _q j  
\end{equation*}
which defines an $n$-valent vertex recursively, for $n>3$, 
\begin{equation*}
   \left\langle  \, \epsfcenter{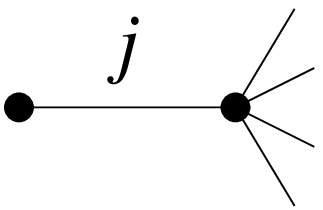}\right\rangle _R=\delta _{j0}
\left\langle \, \epsfcenter{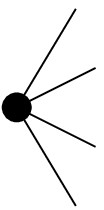} \right\rangle _R 
\end{equation*}
for 1-valent vertices, and
\begin{equation*}
\left\langle \epsfpad{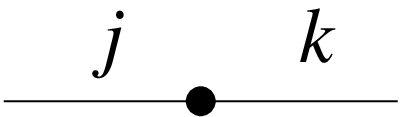}{0pt}{10pt}  \right\rangle _R=\frac{1}{\dim _qj} \delta _{jk}
\left\langle  \epsfpad{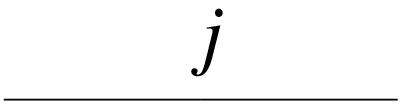}{0pt}{10pt} \right\rangle _R
\end{equation*}
for 2-valent vertices.  

The relation between the state sum invariant of a graph $Z(S^3,\Gamma )$ and
the relativistic invariant $\left\langle \Gamma \right\rangle _R$ is given by a Fourier transform in the spin
labels, using the kernel
\[
K_b(a)=(-1)^{2b} \frac{\sin \frac{\pi}{r} (2a+1)(2b+1)}{\sin \frac{\pi}{r}(2a+1)}.
\]
The result is

\begin{theorem}  
\begin{multline}\label{theorem}
\sum _{{j_1j_2 \dots}  j_n} \frac {Z(S^3,\Gamma(j_1,j_2,\ldots,j_n))} {Z(S^3)} K_{i_1}(j_1)K_{i_2}(j_2)\dots K_{i_n}(j_n)\\
=   \left\langle  \Gamma(i_1,i_2,\ldots,i_n) \right\rangle _R.
\end{multline}
\end{theorem}  

A general proof of this result will appear elsewhere.  However I will prove a particular special case which is interesting, as the result implies some new identities among quantum $6j$-symbols (Appendix 2). This example is also sufficient to provide a proof of the results for the edge and the triangle given earlier.

The example is the tetrahedral graph embedded in $S^3$. The definition of the state sum invariant is
\begin{multline}\label{proof1}
Z\biggl(S^3,
{\def\epsfsize#1#2{0.40#1} 
\epsfpad{tetrahedron.eps}{5pt}{5pt}}
 \biggr)
  \\
=  \frac{ \dim _qj_1 \dots \dim _qj_6} 
{N^4\,\Theta(j_1,j_2,j_3)\Theta(j_1,j_5,j_6)\Theta(j_3,j_4,j_5)\Theta(j_2,j_4,j_6)}
{\def\epsfsize#1#2{0.40#1} 
\Biggl\langle \epsfpad{tetnet.eps}{5pt}{5pt}}
 \Biggr\rangle^2 ,
\end{multline}
since $S^3$ can be `triangulated' with two tetrahedra. The following calculations prove the theorem for this example.

Using Roberts' chain mail~\cite{ROCHM}, the square of the spin network evaluation on the right-hand side can be expressed as a link diagram in which some components are labelled with the formal linear combination 
 $$\Omega=\sum_j(\dim_qj) j$$
 of spins.
 
\begin{multline}\label{proof2}\frac{ 1} 
{\Theta(j_1,j_2,j_3)\Theta(j_1,j_5,j_6)\Theta(j_3,j_4,j_5)\Theta(j_2,j_4,j_6)}
\def\epsfsize#1#2{0.40#1} 
\Biggl\langle \epsfpad{tetnet.eps}{5pt}{5pt}
\Biggr\rangle^2\\
=\frac 1 {N^4}\quad \left\langle\epsfcenter{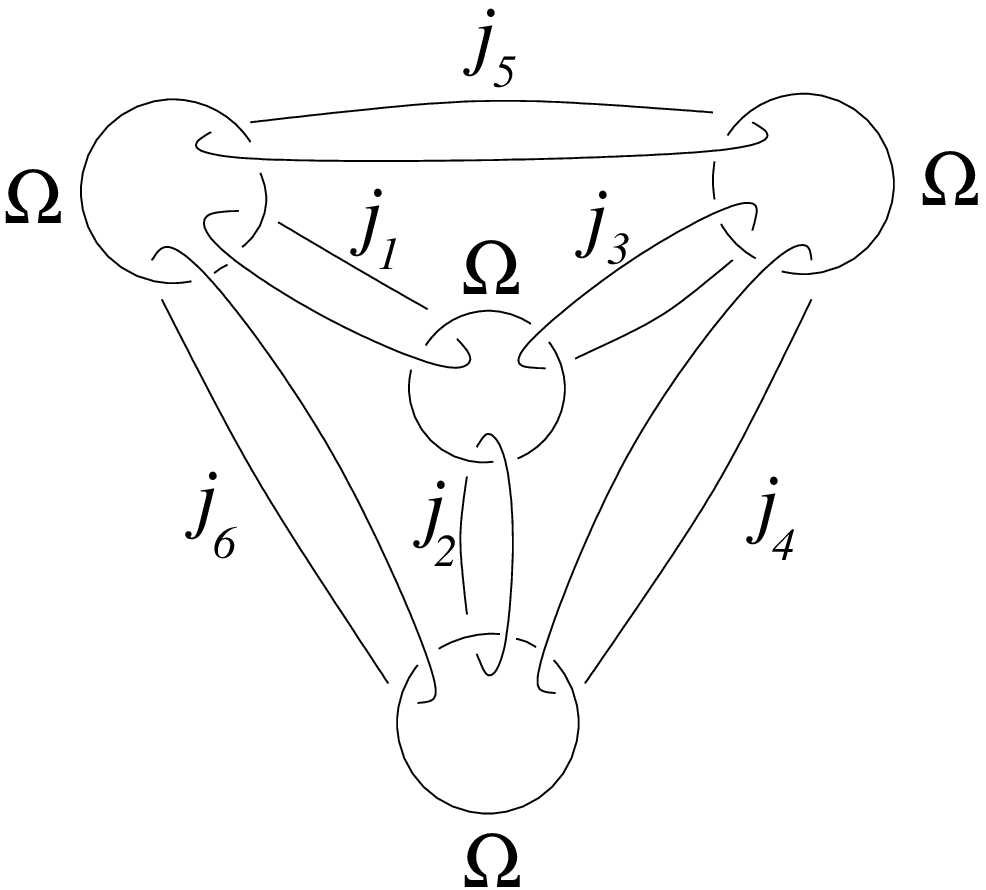}\right\rangle
=\frac 1 {N^3}\quad \left\langle\epsfcenter{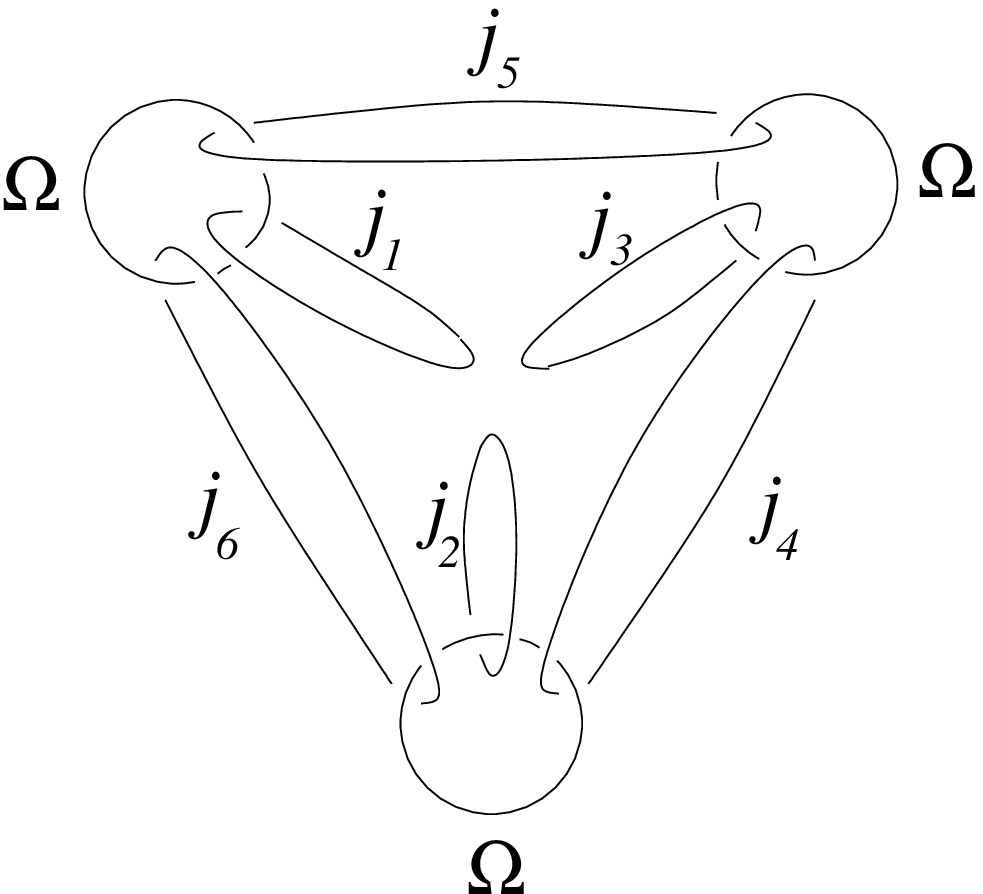}\right\rangle
\end{multline}
using the handleslide identity for $\Omega$.

The Fourier transform kernel is related to the Hopf link
\[
K_i(j)=\left\langle \epsfcenter{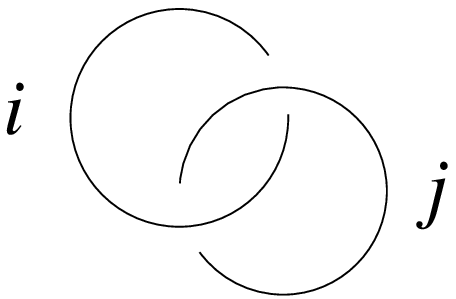} \right\rangle \frac{1}{\dim _q j},
\]
 and the action of the Fourier transform on an edge of a spin network is given by the replacement
 $$\sum_j K_i(j)\dim_q j\quad \epsfpad{line.eps}{0pt}{10pt}\quad=\quad\epsfpad{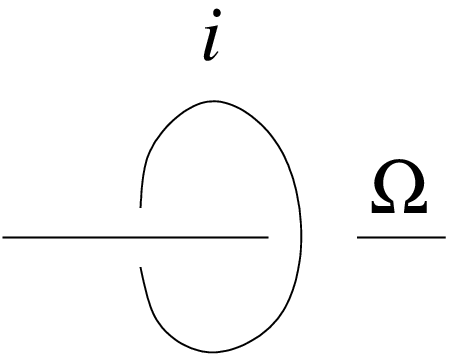}{0pt}{10pt}$$

Applying the Fourier transform to (\ref{proof2}) gives
\begin{multline}\label{proof3}
\sum _{{j_1j_2 \dots}  j_6}  Z\biggl(S^3,
 {\def\epsfsize#1#2{0.40#1} 
\epsfpad{tetrahedron.eps}{5pt}{5pt}}
 \biggr)  
K_{i_1}(j_1)K_{i_2}(j_2)\dots K_{i_6}(j_6)\\
=\frac1{N^7}
\left\langle\epsfcenter{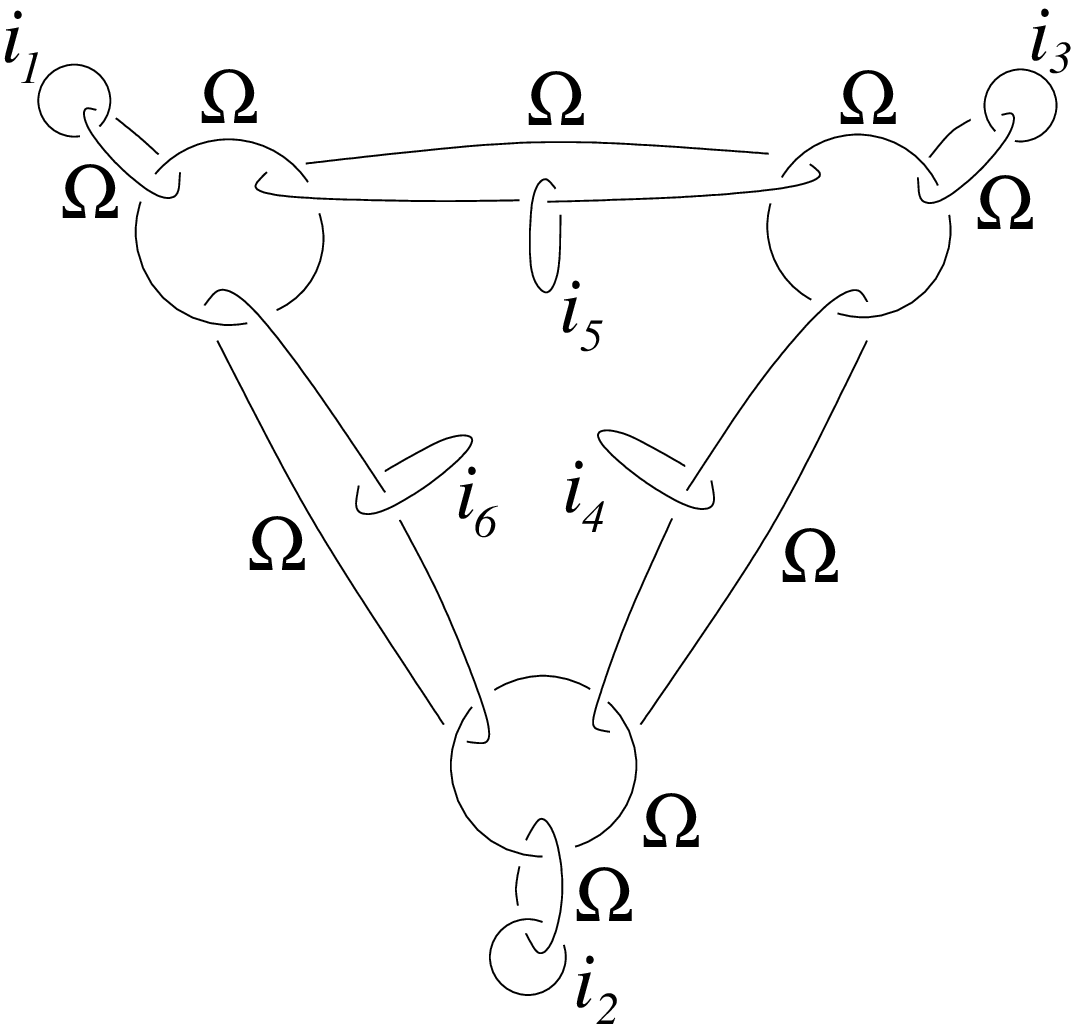}\right\rangle
=\frac1{N^4}
\left\langle\epsfcenter{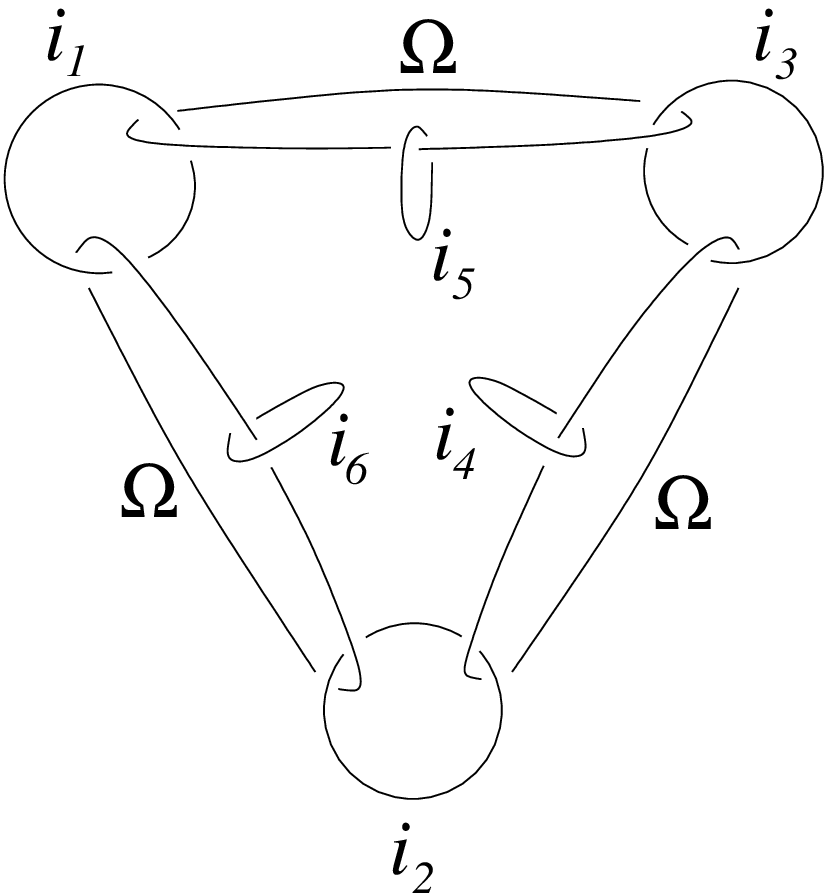}\right\rangle\\
=\frac1{N\,\Theta(i_1,i_2,i_6)\Theta(i_2,i_3,i_4)\Theta(i_1,i_3,i_5)\Theta(i_4,i_5,i_6)} \Biggl\langle
{\def\epsfsize#1#2{0.40#1} 
\epsfcenter{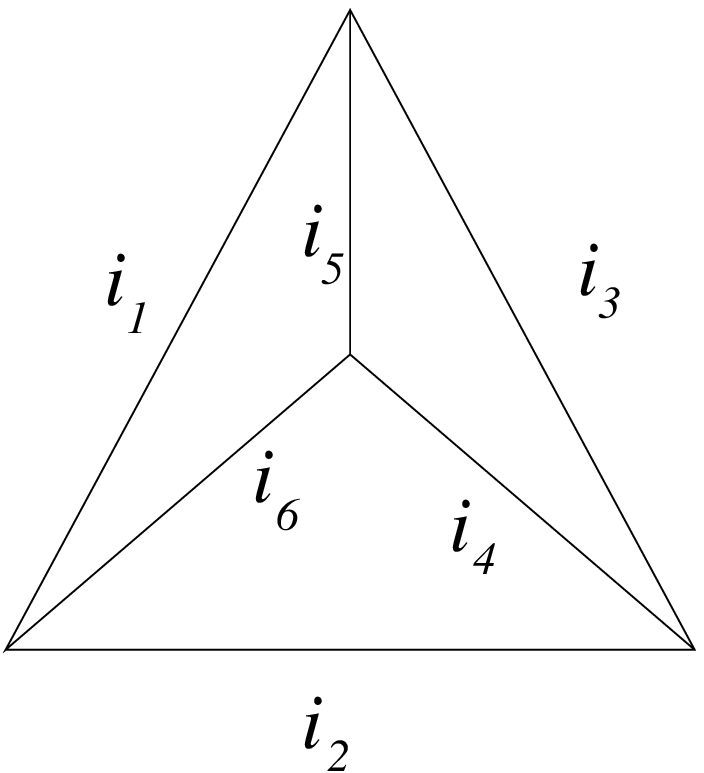}}
\Biggr\rangle^2\\
= {Z(S^3)} \Biggl\langle   
{\def\epsfsize#1#2{0.40#1} 
\epsfcenter{tetneti.eps}}
 \Biggr\rangle _R.
\end{multline}
The graph in the final relativistic spin network is the same as the graph of edges in the original partition function $Z$.
But now the admissibility conditions apply to triples of spins meeting at a vertex of the graph, whereas they applied to triples around a triangular circuit of the original graph in $Z$.

From this example it is possible to prove the theorem very easily also for sub-graphs of the tetrahedron. Setting, for example, $i_1=0$ in (\ref{proof3}) gives, on the left-hand side, a summation over $j_1$ weighted with $K_0(J_1)=1$, which gives the correct state sum formula for the graph with this edge removed, whilst on the right-hand side this gives the relativistic invariant for the graph also with this edge removed. The results at the beginning of the paper can be checked very easily. For example, the relativistic spin network evaluation for $\epsfpad{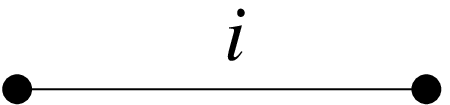}{0pt}{10pt}$ is $\delta_{i0}$ and inverting the transform gives $P_j=\sum_i \dim_q^2j\, K_i(j)\,\delta_{i0}=\dim_q^2j$.

There is a curious analogy between the Fourier transform 
and the duality between position and momentum variables of a particle in quantum theory. In fact the kernel $K_j(a)$ of the Fourier transform is a discrete version of the `zonal spherical function' on $S^3$.
The Laplace operator on $S^3$ (with radius $r/2\pi$) has eigenvalues
\[
\nabla ^2 \phi = \frac{-16\pi ^2}{r^2}j(j+1) \phi
\]
for non-negative half-integer $j$; the eigenfunction that is spherically symmetric about $p\in S^3$ (the zonal spherical function) is
\[
G_j(R)=(-1)^{2j}\frac{\sin \frac{2\pi}{r}(2j+1)R}{\sin \frac{2\pi}{r}R},
\]
where $R$ is the distance from $p$.

Putting $R=a+\frac{1}{2}$ shows that at half-integer values, $G$ coincides with the Fourier transform kernel 
\[ K_j(a)=G_j(a+\frac{1}{2}) \]
so that the Fourier transform can be interpreted as a transition to a
sort of `momentum' or `mass' representation for the quantum
probabilities.

\section*{Appendix 1. 3 points on $S^3$}

If three points are distributed on $S^3$ with uniform probability, then this determines a probability distribution on the space of distances between these three points.

The 3-sphere has standard spherical coordinates $(\chi,\theta,\phi)$ which determine points in $S^3\subset \R^4$ by
$$
\frac r {2\pi} \left(\cos\chi,\sin\chi\cos\theta,\sin\chi\sin\theta\cos\phi,\sin\chi\sin\theta\sin\phi\right).
$$
Using the rotational symmetry, three points on $S^3$ can be assumed to be at $(\chi,\theta,\phi)$ coordinates
\begin{align*}
p&=(0,0,0)\\
q&=(\chi_2,0,0)\\
s&=(\chi_3,\theta_3,0)
\end{align*}
The probabilty is thus
$$ \rd P= \frac 2\pi \sin^2\chi_2\,\rd\chi_2 \,\frac 1\pi \sin^2\chi_3\,\sin\theta_3\, \rd\chi_3\, \rd\theta_3.$$
For three points on a 3-sphere of radius $r/2\pi$, the distances between them (\fig{proof}) are given by
\begin{align*}
R_2&=\frac r {2\pi}\chi_2\\
R_3&=\frac r {2\pi}\chi_3\\
\cos\frac{2\pi}r R_1 &= \cos\chi_2\cos\chi_3+ \sin\chi_2\sin\chi_3\cos\theta_3,
\end{align*}
the last equation being the cosine law for the spherical triangle $pqs$ with $\theta_3$ the angle at $p$.

Differentiating these relations gives
$$
\rd P=\frac {16\pi}{r^3} \sin\frac{2\pi R_1}r \, \sin\frac{2\pi R_2}r \, \sin\frac{2\pi R_3}r  \,\rd R_1\, \rd R_2\, \rd R_3
$$
when the inequalities for a spherical triangle are satisfied, and zero otherwise.

\section*{Appendix 2. Identity for 6$j$-symbols}

The 6$j$-symbols are defined to be normalised versions of the tetrahedral spin network evaluation~\cite{KR}:
\begin{multline*}
\left\{\begin{matrix}j_1 & j_2 & j_3\\ j_4 & j_5 & j_6\end{matrix}\right\}_q=\\
 \frac 1 {\sqrt{ \Theta(j_1,j_2,j_3) \Theta(j_1,j_5,j_6)
\Theta(j_3,j_4,j_5)
\Theta(j_2,j_4,j_6)} }
\left<\epsfpad{tetnet.eps}{5pt}{5pt}\right>.
\end{multline*}
Using this definition, the identity proved after the statement of the theorem is
\begin{equation*}
\frac 1{N^3}
\sum_{j_1\ldots j_6}
\left\{\begin{matrix}j_4 & j_5 & j_6\\ j_1 & j_2 & j_3\end{matrix}\right\}_q^2
H(j_1,i_1)\ldots H(j_6,i_6)
=\left\{\begin{matrix}i_1 & i_2 & i_3\\ i_4 & i_5 & i_6\end{matrix}\right\}_q^2
\end{equation*}
where
\begin{equation*}
H(j,i) = K_i(j)\dim_qj= \frac{\sin\frac\pi r(2i+1)(2j+1)}{\sin\frac\pi r}(-1)^{2i+2j}.
\end{equation*}
The identity does not appear to have a classical ($q=1$) analogue.


\begin{thebibliography}{99}
\bibitem{QGTQFT}J. W. Barrett,  Quantum gravity as topological quantum field theory. J.~Math.\ Phys. {\bf 36}  6161--6179 (1995)

\bibitem{CLEV} J.W.\ Barrett, The classical evaluation of relativistic spin
networks. Advances in Theoretical and Mathematical Physics {\bf 2} 593--600 (1998)

\bibitem{HK} J.-C.\ Hausmann and A.\ Knutson, Polygon spaces and 
Grassmannians, {\sl L'Enseignment Mathematique} {\bf 43} (1997), 173-198.   

\bibitem{KM} M.\ Kapovich and J.\ Millson, The symplectic geometry of 
polygons in Euclidean space, {\sl Jour.\ Diff.\ Geom.\ }{\bf 44} 
(1996) 479-51. 

\bibitem{KS} M. Karowski and R. Schrader, A combinatorial approach to topological quantum field theories and invariants of graphs. Commun.\ Math.\ Phys. {\bf 151} 355--402 (1993)

\bibitem{KB}  L.H. Kauffman and S.L. Lins, Temperley-Lieb recoupling theory and invariants of 3-manifolds. Princeton UP   (1994)
 
\bibitem{KR} A.N. Kirillov, N.Yu.\ Reshetikhin, Representations of the algebra $U_q(sl(2))$, $q$-orthogonal polynomials and invariants of links.
 In: Infinite-Dimensional Lie Algebras and Groups. Ed.\ V.G. Kac, World Scientific. 285--339 (1989)

\bibitem{PR} G.\ Ponzano and T.\ Regge, Semiclassical limit of
Racah coefficients, in {\sl Spectroscopic and Group Theoretical
Methods in Physics,} ed.\ F.\ Bloch, North-Holland, New York, 1968.

\bibitem{ROCHM} J. Roberts, Skein theory and Turaev-Viro invariants. Topology  {\bf 34} 771--788   (1995)

\bibitem{ROEX}J. Roberts, Refined state-sum invariants of $3$- and $4$-manifolds. Geometric topology (Athens, GA, 1993), 217--234, AMS/IP Stud.
Adv. Math., 2.1, Amer. Math. Soc., Providence, RI. (1997)

\bibitem{TV}V.\ Turaev and O.\ Viro, State sum invariants of 3-manifolds
and quantum $6j$ symbols, {\sl Topology} {\bf 31} (1992), 865-902.

\bibitem{TBIG} V. Turaev, Quantum invariants of links and 3-valent graphs in 3-manifolds. Inst.\ Hautes Etudes Sci.\ Publ.\ Math. {\bf 77} 121--171 (1993)

\bibitem{YEX}D.N. Yetter, Homologically Twisted Invariants Related to (2+1)- and (3+1)-Dimensional State-Sum Topological
Quantum Field Theories. Eprint hep-th/9311082 (1993)

\bibitem{WI3D}E. Witten, 2+1 gravity as an exactly soluble system. Nucl.~Phys.~{\bf B311}  46--78 (1988)

\bibitem{WITCA}E. Witten, Topology-changing amplitudes in 2+1 dimensional gravity. Nucl.~Phys.~{\bf B323}  113--140 (1989)

\bibitem{YGR}D.N. Yetter, Generalized Barrett-Crane vertices and invariants of embedded graphs. J.~Knot Theor.\ Ram.\ {\bf 8} 815--829 (1999). 

\end{thebibliography}
\end{document}